\documentclass[aps,prl,citeautoscript,reprint,showpacs,superscriptaddress,twocolumn,longbibliography]{revtex4-2}

\usepackage{preamble}

\begin{document}

\date{\today}
\title{Power-law entanglement and Hilbert space fragmentation in non-reciprocal quantum circuits}

\author{K. Klocke}
\affiliation{Department of Physics, University of California, Berkeley, California 94720, USA}
\author{J. E. Moore}
\affiliation{Department of Physics, University of California, Berkeley, California 94720, USA}
\affiliation{Materials Sciences Division, Lawrence Berkeley National Laboratory, Berkeley, California 94720, USA}
\author{M. Buchhold}
\affiliation{Institut f\"ur Theoretische Physik, Universit\"at zu K\"oln, D-50937 Cologne, Germany}


\begin{abstract}
Quantum circuits utilizing measurement to evolve a quantum wave function offer a new and rich playground to engineer unconventional entanglement dynamics. 
Here we introduce a hybrid, non-reciprocal setup featuring a quantum circuit, whose updates are conditioned on the state of a classical dynamical agent. 
In our example the circuit is represented by a Majorana quantum chain controlled by a classical $N$-state Potts chain undergoing pair-flips. 
The local orientation of the classical spins controls whether randomly drawn local measurements on the quantum chain are allowed or not. 
This imposes a dynamical kinetic constraint on the entanglement growth, described by the transfer matrix of an $N$-colored loop model. 
It yields an equivalent description of the circuit by an $SU(N)$-symmetric Temperley-Lieb Hamiltonian or by a kinetically constrained surface growth model for an $N$-component height field. 
For $N=2$, we find a diffusive growth of the half-chain entanglement towards a stationary profile $S(L)\sim L^{1/2}$ for $L$ sites. 
For $N\ge3$, the kinetic constraints impose Hilbert space fragmentation, yielding subdiffusive growth towards $S(L)\sim L^{0.57}$. 
This showcases how the control by a classical dynamical agent can enrich the entanglement dynamics in quantum circuits, paving a route toward novel entanglement dynamics in non-reciprocal hybrid circuit architectures.
\end{abstract}

\maketitle


\paragraph{Introduction} --- 
Analog quantum circuits, incorporating unitary gates, mid-circuit measurements, and quantum feedback, extend beyond traditional Hamiltonian or dissipative quantum systems. 
By utilizing measurement as a resource for irreversible quantum dynamics, quantum circuits offer a unique platform for shaping and manipulating entanglement. 
They enable applications such as quantum error correction~\cite{Stricker_2020, Ryan_Anderson_2021, Erhard_2021, Krinner_2022, Livingston_2022, Bluvstein_2023, Norcia_2023, Ma_2023, Singh_2023}, expedited state preparation, and the exploration of novel phenomena like measurement-induced entanglement transitions~\cite{Li_2019, Skinner_2019, Choi_2020, Bao_2020, Buchhold_2021, Li_2021_statistical, Li_2021_hybrid, Ippoliti_2021, Sang_2021_PRR, Sang_2021_PRXQ, Muller_2022, Noel_2022, Fisher_2023_review, Chertkov_2023}.
In particular, the efficient preparation of entanglement patterns associated to circuit analogues of topologically nontrivial or gapless ground states has recently attracted attention due to the challenge of preparing such states by exclusively using unitary operations~\cite{Piroli_2021, Bravyi_2022, Tantivasadakarn_2022, Verresen_2022, Lu_2022, FossFeig_2023, Iqbal_2023, Tantivasadakarn_2023_PRXQ, Tantivasadakarn_2023_PRL, Lu_2023, Iqbal_2024}.

Here we propose a novel route to implement enriched entanglement dynamics -- the non-reciprocal coupling of a quantum circuit to a classical dynamical agent. Non-reciprocity indicates that the evolution of the quantum circuit is influenced by the classical agent, but not vice versa.
The dynamical agent evolves according to a set of predetermined update rules and typically implements a non-Markovian environment, which governs the build-up of entanglement. 
We consider a system composed of a Majorana quantum circuit conditioned on a classical spin chain undergoing pair flip dynamics. 
Utilizing the loop model framework for Majorana circuits~\cite{Klocke_2023, Nahum_2020}, we show that this circuit features both a dynamical fragmentation of the Hilbert space and an algebraic growth of the entanglement entropy $S(L) \sim L^{\alpha}$ familiar from $SU(N)$-symmetric quantum models such as, e.g., the Fredkin chain~\cite{Bravyi_2012, Salberger_2016, Movassagh_2016, Movassagh_2017}. 
We validate this picture by extensive numerical simulations of the non-reciprocal circuit.

\begin{figure}[!t]
    \centering
    \includegraphics[width=\columnwidth]{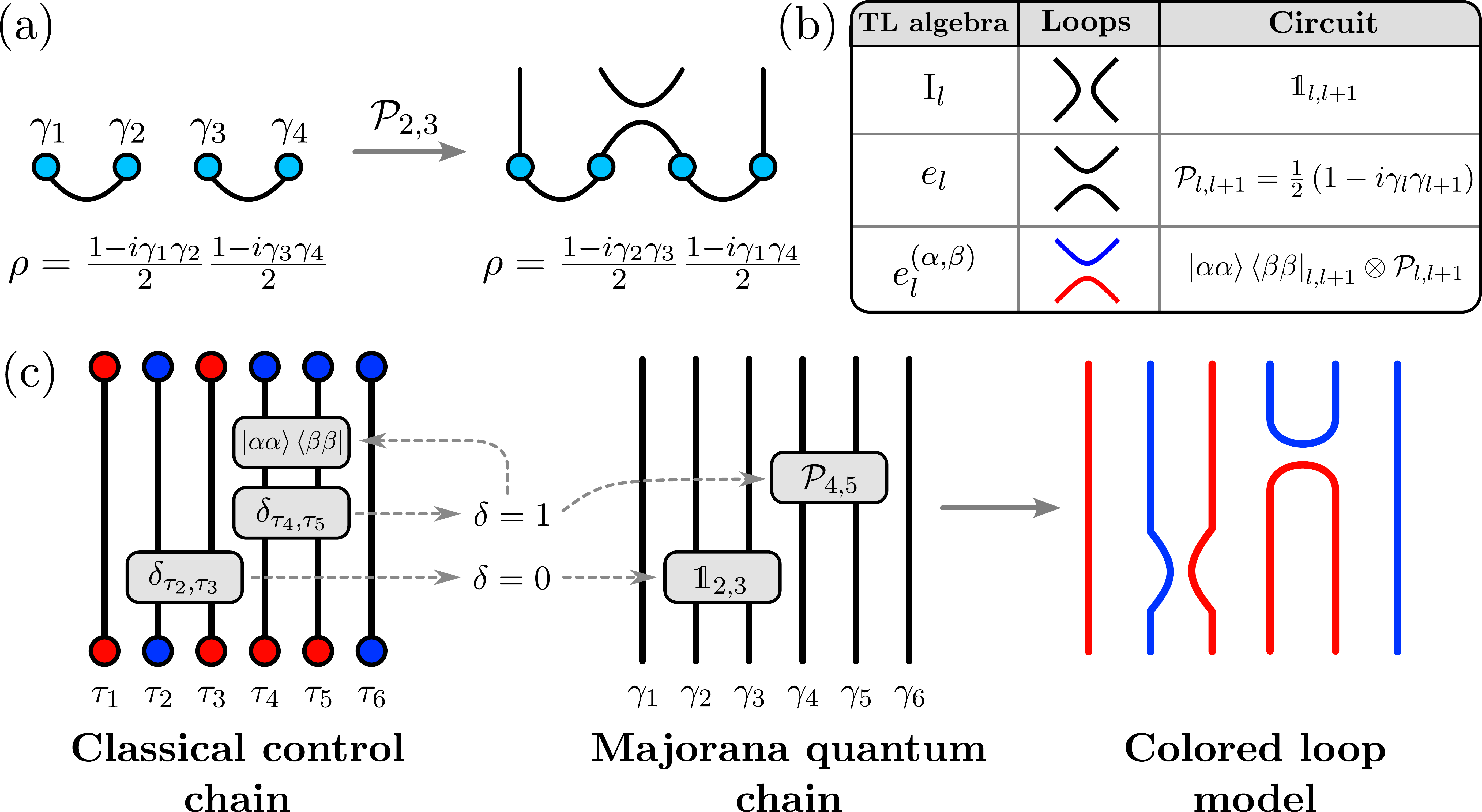}
    \caption{
    \textbf{Non-reciprocal circuit for colored loop models.}
    (a) Diagrammatic representation of Majorana worldlines being rewired under projective measurement.
    (b) A dictionary relating the TL algebra elements to their loop representation and corresponding circuit operation.
    (c) Effective non-reciprocal circuit for the colored loop model. If and only-if a brick $\{\tau_l,\tau_{l+1}, \gamma_l, \gamma_{l+1}\}$ has two identical Potts spins $\tau_l=\tau_{l+1}$ an update is performed: the parity $i\gamma_l\gamma_{l+1}$ is measured and the Potts spins are randomly flipped $\tau_l,\tau_{l+1}\to \beta$.
    }
    \label{fig:model_summary}
\end{figure}

\paragraph{Majorana loop model circuits.} ---
We consider the dynamics of entanglement in  Clifford circuits of $L$ Majorana fermions $\{\gamma_l,\gamma_m\}=2\delta_{l,m}$ on a one-dimensional chain with open boundary conditions (OBC).
The evolution is generated by parity checks of bilinears $i\gamma_l\gamma_m$, yielding a stabilizer state $\rho = \prod_\alpha \frac{\mathbbm{1} + i\gamma_{l_\alpha}\gamma_{m_\alpha}}{2}$ with well-known parities $i\gamma_{l_\alpha}\gamma_{m_\alpha}$.
The state $\rho$ admits a graphical representation via Majorana worldlines, each of which connects two Majorana fermions $\gamma_l,\gamma_m$ with well-defined parity $i\gamma_l\gamma_m=\pm1$~\cite{Klocke_2023, Sang_2021_PRXQ, Sang_2021_PRR, Nahum_2020, Merritt_2023}. 
Performing a parity check $i\gamma_l\gamma_m=\pm 1$ acts with either the projector $\calP_{l,m} = \tfrac12(\mathds{1} - i\gamma_l\gamma_m)$ or its sign-flipped partner $\calP_{m,l}$ on $\rho$, thus rewiring the worldlines (see Fig.~\ref{fig:model_summary}a).
The entanglement entropy of a subsystem $A$, $S_A = -\Tr[\rho_A \log_2(\rho_A)]$  is agnostic to the parity sign and only depends on the modulus $\abs{\langle i\gamma_l\gamma_m\rangle}$.
This has important consequences: both $\calP_{l,m}, \calP_{m,l}$ implement precisely the same change in entanglement, defining an equivalence relation for projections $\calP_{l,m} \sim \calP_{m,l}$ such that circuits are elements of the Brauer algebra $\mathcal{B}_L(1)$ over $L$ Majorana fermions~\cite{Klocke_2023}.

For nearest-neighbor Majorana measurements, the fundamental generators of dynamics are projectors $e_{l}\equiv\calP_{l,l+1}$. They obey the Temperley-Lieb (TL) algebra~\footnote{Including the normalization of the wave function after each projection.}
\be
    e_{l}^2 = n e_{l}, \quad e_{l}e_{l\pm1}e_{l} = e_{l},
    \label{eq:TL_algebra}
\ee
describing loop models with fugacity $n$.
Projective measurements exclusively implement $n=1$. 
Depending on geometry and the set of measured operators, Majorana circuits with $n=1$ realize a spectrum of phases and critical points~\cite{Klocke_2023, Nahum_2020, Lang_2020, Li_2023_Z2, Roser_2023}, e.g., of (coupled) Potts models. 


\paragraph{Colored loop models from non-reciprocal circuits.} --- 
In order to further enrich the quantum dynamics in Majorana circuits and to reach beyond the $n=1$ paradigm, we introduce a non-reciprocal circuit architecture implementing a \emph{colored} TL algebra~\cite{Grimm_1993, Grimm_2003, Grimm_1995}. 
This algebra is generated by projectors $e_l^{(\alpha,\beta)}$ with fugacity $n=1$ that are additionally decorated with $N$ color degrees of freedom, $\alpha,\beta$.
The operators obey the algebra~\footnote{Technically speaking there is also a second relation like in Eq.~\eqref{eq:TL_algebra} corresponding to ambient isotopy of colored loops. It is $e_l^{(\alpha,\nu)} e_{l\pm 1}^{(\nu,\nu)} e_l^{(\nu,\beta)} = e_l^{(\alpha,\beta)} P_{l\pm 1}^\nu$, where $P_l^\nu$ is a projector on the color degree of freedom on the corresponding worldline.}
\be
    e_l^{(\alpha,\nu)}e_l^{(\nu',\beta)} = \delta_{\nu, \nu'}e_l^{(\alpha,\beta)}. 
    \label{eq:color_algebra}
\ee

In the loop model framework, $e_l^{(\alpha,\beta)}$ closes a loop of color $\beta$ and opens a new loop of color $\alpha$, see Fig.~\ref{fig:model_summary}a.
Averaging over all colors with equal probability yields the generators $E_l = \sum_{\alpha,\beta = 1}^N e_l^{(\alpha,\beta)}$, which implement the (uncolored) TL algebra with fugacity $N$~\cite{Nahum_2013_3D, Kaul_2013, Evertz_2003},
\be
    E_l^2=NE_l, \quad E_l E_{l\pm1}E_l=E_l.
    \label{eq:TL_alg_N}
\ee
Such a colored TL algebra arises, e.g., in the pair-flip model~\cite{Caha_2018, Han_2024}, where the color symmetry imposes kinetic constraints resulting in Hilbert space fragmentation and unconventional entanglement scaling in the ground-state.

In order to implement the algebra in Eq.~\eqref{eq:color_algebra}, we augment the quantum circuit of $L$ Majorana fermions with a ``control'' chain of $L$ classical $N$-state Potts spins $\tau_l \in \{1,\dots,N\}$.
The classical spins influence the evolution of the Majorana chain via a non-reciprocal interaction: At each time step $t$, a ``brick'' $\{\gamma_l, \gamma_{l+1}, \tau_l, \tau_{l+1}\}$ consisting of two adjacent Majorana fermions and two corresponding Potts spins is chosen randomly from all $L-1$ possible pairings (OBC). 
\emph{If and only-if} the two Potts spins have the same ``color'' ($\tau_l = \tau_{l+1} = \alpha$), two updates are performed: 
(i) the parity operator $i\gamma_l\gamma_{l+1}$ is measured, 
(ii) both Potts spins are flipped to a randomly and uniformly chosen color $\tau_l,\tau_{l+1}\to\beta\in \{1,...,N\}$. 
This implements the generator of the colored TL algebra
\be
    e_l^{(\alpha,\beta)} = \ket{\alpha\alpha}\bra{\beta\beta}_{l,l+1} \otimes \calP_{l,l+1}.
    \label{eq:colored_projector}
\ee
The Potts spins thus decorate the Majorana worldlines, yielding a colored loop model for the effective dynamics.


\paragraph{Ensemble Hamiltonian.} --- Let $\epsilon_{\alpha\beta}^{(l)}\in[0,1]$ be the probability applying $e^{(\alpha,\beta)}_l$ site $l$ in a unit time step. Averaging over all circuit trajectories then reproduces a loop ensemble generated by an effective transfer matrix $T_l$ acting on each pair of neighboring sites as $T_l = \mathds{1} + \epsilon_{\alpha\beta}^{(l)}e^{(\alpha,\beta)}_l$.
This is exactly the transfer matrix arising from imaginary-time evolution with a Hamiltonian 
\be
    H = \sum_l \sum_{\alpha,\beta} \epsilon_{\alpha\beta}^{(l)} e_{l}^{(\alpha,\beta)}.
    \label{eq:Ham}
\ee
Drawing the probabilities uniformly in space $\epsilon_{\alpha\beta}^{(l)}\equiv \epsilon_{\alpha\beta}$, Eq.~\eqref{eq:Ham} is the so-called pair-flip (PF) Hamiltonian. 
Choosing further identical transition rates $\epsilon_{\alpha\beta}\equiv \epsilon$ yields the TL Hamiltonian $H = \epsilon\sum_l E_l$. 
The transfer matrix picture elucidates that each trajectory of the random circuit represents an element of the stochastic series expansion for the partition function $\sim \exp(-\beta H)$ in the limit $\beta\to\infty$.
The ensemble Hamiltonian thus is the Markov generator for a classical stochastic process which produces individual circuit trajectories (i.e. colored loop configurations)~\cite{Pearce_2002, DeGier_2003, ZinnJustin_2009}.
The statistical ensemble of wave functions produced by the circuit is identical to the ensemble that forms the quantum ground state of the  Hamiltonian~\eqref{eq:Ham} by coherent superposition.


\paragraph{Symmetries and Fragmentation.} ---  
The TL generators $e_l^{(\alpha,\beta)}$ induce pair-flip dynamics that conserve $N-1$ independent $U(1)$ charges $Q_\alpha = \sum_l (-1)^l \ket{\alpha}\bra{\alpha}_l$~\cite{Hart_2023, Moudgalya_2021}. 
Their presence implies a continuity equation yielding diffusive charge transport.
The TL Hamiltonian maps exactly to the spin-$1/2$ Heisenberg chain for $N=2$ and the spin-1 biquadratic chain $H_{N=3} = -\sum_l \left(\vec{S}_l \cdot \vec{S}_{l+1}\right)^2$ for $N=3$~\cite{Klümper_1989, Parkinson_1988, Barber_1989, Chen_2017}, and the $U(1)^{N-1}$ symmetry is promoted to $SU(N)$.
We note, however, that for the circuit, $SU(N)$ is a symmetry of the ensemble, not of individual trajectories.
In addition to the global symmetries, there is an extensive number of additional conserved quantities, which for $N=2$ can be attributed to integrability.

For $N \geq 3$, the TL and PF models are paradigmatic examples of strong Hilbert space fragmentation~\cite{Moudgalya_2021, Han_2024}, with the largest Krylov sector comprising a measure-zero subset of the Hilbert space in the thermodynamic limit~\cite{Moudgalya_2021, Caha_2018}.
This separates $N\ge3$ from $N=2$: $N=2$ has a commutant algebra of dimension $L+1$ and thus fragmentation is absent. $N \geq 3$ gives an exponentially large commutant algebra of dimension $((N-1)^{L+1} - 1)/(N-2)$. 
We thus expect transport and the buildup of entanglement to be slower (subdiffusive) for $N\ge3$ compared to diffusion for $N=2$.

A note on the nature of fragmentation: While the PF model with $N \geq 3$ displays fragmentation in a local product basis --  referred to as ``classically'' fragmented -- the TL model displays fragmentation in an entangled basis -- referred to as ``quantum'' fragmented. 
In our setup, the Potts spins undergo a stochastic classical pair-flip dynamics, with the colors always in a product state.
Even though the ensemble retains a statistical $SU(N)$ symmetry, the trajectory-level dynamics ought to reflect the classical fragmentation of the PF model. 
This fragmentation is imprinted upon the entangled quantum dynamics of the Majorana chain via the non-reciprocal interaction.


\paragraph{Surface growth models and universality.} --- 
Besides the link to fragmented quantum Hamiltonians, the dynamics of Majorana worldlines is directly connected to the dynamics of height models describing surface growth. 
This grants access to dynamical scaling exponents governing the growth of entanglement in the circuit. 

\begin{figure}[!t]
    \centering
    \includegraphics[width=\columnwidth]{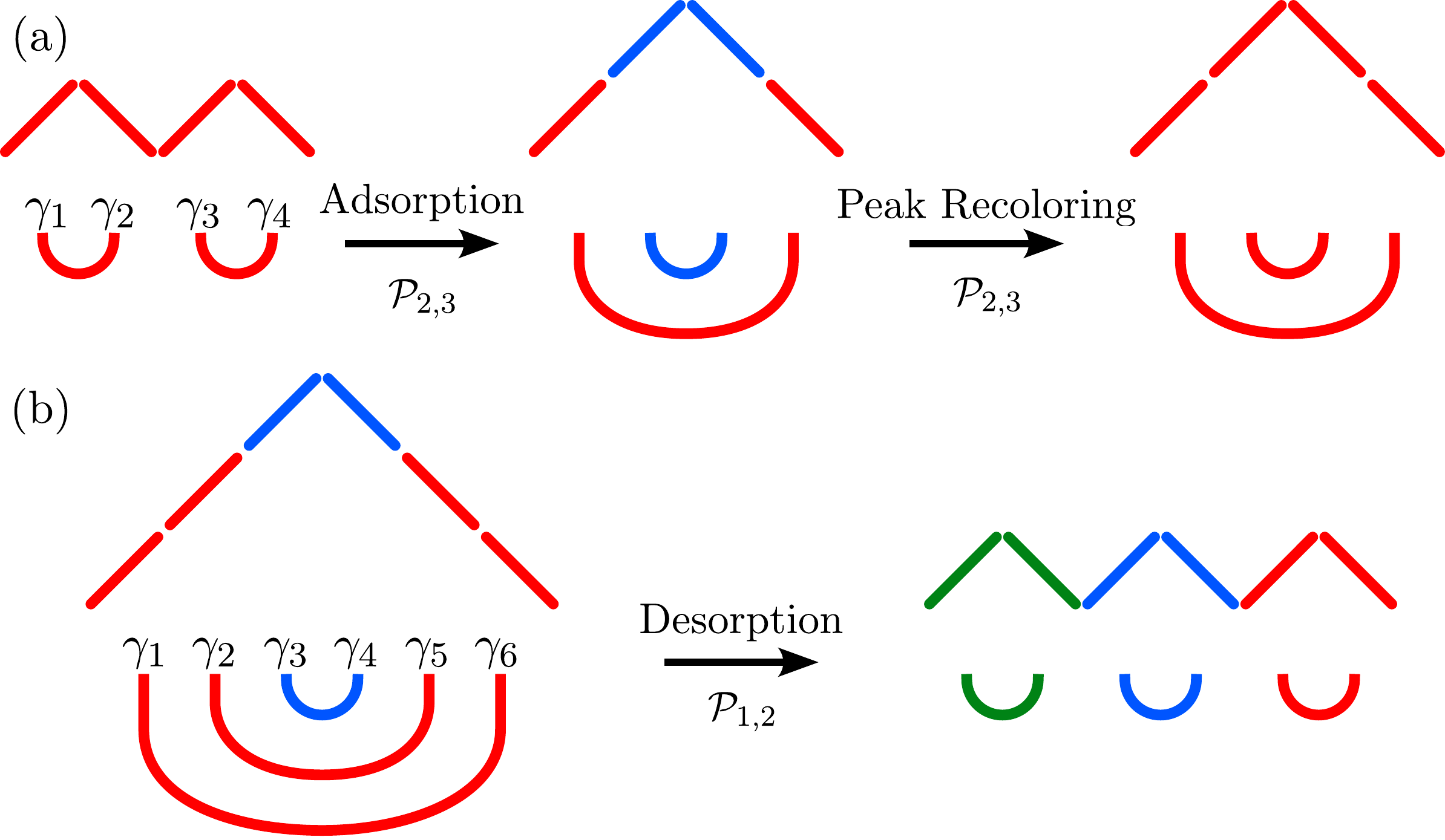}
    \caption{
    \textbf{Height model and Dyck path representation of the colored circuit.}
    Given a loop configuration, the corresponding Dyck path is defined by increasing (decreasing) the path height at the left (right) end of a loop.
    Measurement implements one of the following operations on the height field.
    (a) At a local minimum, measurement implements single-site adsorption, wherein the local height increases.
    At a local maximum, measurement can recolor the peak of the path.
    (b) Otherwise, measurements trigger a desorption avalanche, wherein the height is decreased in the whole region between the endpoints of the loop acted upon.
    }
    \label{fig:dyck_path_mapping}
\end{figure}

Consider a stabilizer state and its loop configuration in the uncolored Majorana chain. 
We define a height field $h_l = S_{[1,l]}$ as the entanglement entropy of the first $l$ Majorana fermions. 
The height field satisfies the boundary conditions $h_0 = h_L = 0$ and the restricted solid-on-solid (RSOS) constraint $h_{l+1} - h_l = \pm 1$. 
Edges connecting adjacent heights then define a Dyck path~\cite{DeGier_2003, ZinnJustin_2009} (i.e., one with $h_l \geq 0$), for which left (right) ends of loops correspond to an increase (decrease) in the height, see Fig.~\ref{fig:dyck_path_mapping}. 

A projective measurement $\calP_{l, l+1}$ then implements one of the following two processes:
(i) single-site adsorption (Fig.~\ref{fig:dyck_path_mapping}a) or (ii) a desorption avalanche (Fig.~\ref{fig:dyck_path_mapping}b).
Adsorption converts a local minimum $\{\dots, h, h-1, h, \dots\}$ to a local maximum $\{\dots, h, h+1, h, \dots\}$, increasing entanglement locally across a cut.
Desorption occurs when measurement acts on a region of constant slope in the height field (e.g. $\{\dots, h-1, h, h+1, \dots\}$), corresponding to two nested loops.
Desorption is disentangling and may cause a non-local ``avalanche'' wherein the whole region enclosed by the inner loop has its height reduced by two.

The colored loop model in the non-reciprocal circuit then yields colored Dyck paths, where the edge connecting $h_l$ and $h_{l+1}$ has the color given by $\tau_l$.
This coloring imposes a kinetic constraint such that adsorption and desorption can occur only when adjacent edges in the Dyck path have the same color.
Moreover, we now have a third process wherein local maxima can be recolored without changing the heights (see Fig.~\ref{fig:dyck_path_mapping}a).

Surface growth models describing a height field $h_l$ give rise to Family-Vicsek (FV) scaling $f(L, t) = L^\alpha F(tL^{-z})$ for quantities $f(L,t)$ related to surface roughness~\cite{Vicsek_1984, Family_1985}.
The scaling function $F(x)$ grows algebraically as $x^\beta$ for small argument and approaches a constant for $x \gg 1$.
The growth exponent $\beta$ is related to the roughness exponent $\alpha$ and the dynamical exponent $z$ via $\beta = \alpha / z$. The relation to surface growth provides lower bounds on the height field $h_l$, i.e., the entanglement growth, and estimates for the universal scaling exponents. 


\paragraph{Entanglement bounds from surface growth.} --- 
The color dynamics on the control chain is reversible, yielding detailed balance and a uniform distribution over states in the accessible Krylov sector. 
The largest Krylov sector in the classical chain yields Dyck paths with the average height growing as $\sqrt{L}$ for all $N$~\cite{Caha_2018}. 
However, for a purely classical chain, i.e., without the Majorana worldlines in the quantum chain, the mapping between Dyck paths and color configurations is not unique. 
For example, the color configuration $\ket{1111}$ might arise in two different loop configurations shown in Fig.~\ref{fig:dyck_path_mapping}a. 
In Ref.~\cite{Caha_2018} each color configuration is mapped to the Dyck path with the smallest possible height, providing a lower bound of $\sqrt{L}$ for the height field and the entanglement entropy.


\paragraph{Critical exponents from surface growth.} --- 
For $N=2$, the TL Hamiltonian~\eqref{eq:Ham} corresponds to the spin-$1/2$ Heisenberg chain, which is a Markov generator for the symmetric simple exclusion process (SSEP)~\cite{Kandel_1990, Gwa_1992_PRL, Doochul_1995, Neergaard_1995}. 
The dynamics of the SSEP lie in the Edwards-Wilkinson (EW) universality class, with exponents $z=2$, $\alpha=\tfrac12$ and $\beta=\tfrac14$~\cite{Edwards_Wilkinson_1982, Kandel_1990, Gwa_1992_PRA, Gwa_1992_PRL, Doochul_1995} \footnote{By contrast, the asymmetric simple exclusion process (ASEP) exhibits KPZ dynamics.
Tuning away from the symmetric point requires conditioning the measurement rates on the local height profile, akin to the unitary dynamics in Ref.~\cite{Morral_2024}.
However, the local Potts configuration does \emph{not} confer information about the height field, thereby protecting the EW universality expected for our circuit with $N=2$.}. 
For $N \geq 3$, Hilbert space fragmentation will modify the critical exponents compared to $N=2$. 
The dynamics of the Hamiltonian~\eqref{eq:Ham} with $N=3$ has been explored in the context of classical deposition-evaporation models for colored dimers and trimers, where numerical results show subdiffusion with exponent $z \approx 2.5$ and growth exponent $\beta \approx 0.19$~\cite{Koduvely_1998, Menon_1995, Menon_1997, Barma_1994}.
More recent studies on dynamics in the Fredkin chain, which admits a closely related description in terms of kinetically constrained surface growth, yield a dynamical exponent $z \approx 2.7$ in the zero magnetization sector~\cite{Singh_2021, Chen_2017, Adhikari_2021}.


\paragraph{Dynamics of entanglement.} --- 
We simulate the growth of entanglement in the non-reciprocal circuit, starting from an initial state with uniform coloring $\tau_l = 1$ and local pairing $\langle i \gamma_{2l-1}\gamma_{2l} \rangle = 1$.
This corresponds to a ``flat'' height configuration, lying in the largest Krylov sector, where each loop has a well-defined color.

\begin{figure}[!t]
    \centering
    \includegraphics[width=\columnwidth]{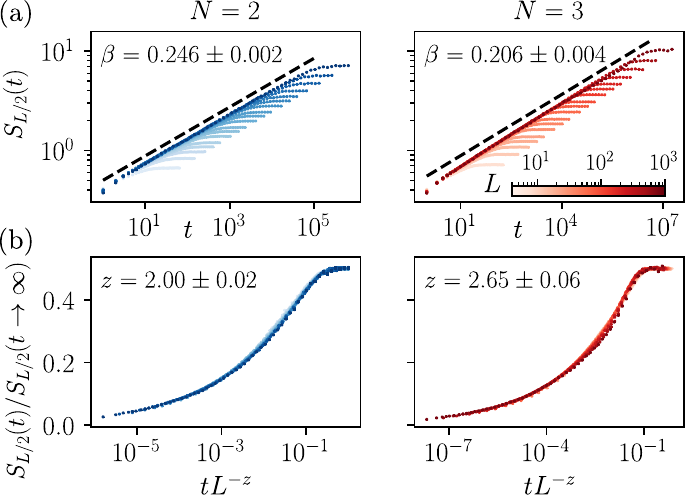}
    \caption{
    Growth of the half-chain entanglement entropy beginning from a trivial initial state for $N=2$ (left) and $N=3$ (right).
    (a) We observe an algebraic scaling $S_{L/2}(t) \sim t^{\beta_N}$, where the growth exponent is $\beta_2 = 0.246 \pm 0.002$ and $\beta_3 = 0.206 \pm 0.004$.
    (b) Rescaling of time yields a data collapse of the normalized entanglement entropy $S_{L/2}(t) / S_{L/2}(t\rightarrow\infty)$.
    For $N=2$ we find dynamical exponent $z = 2.00 \pm 0.02$, while for $N=3$ we find $z = 2.65 \pm 0.06$.
    }
    \label{fig:dynamics}
\end{figure}

We provide detailed simulation results of both $N=2$ and $N=3$ colors. 
Here $N=3$ is a representative for generic $N\ge3$, while $N=2$ is special since it has distinct symmetries and no Hilbert space fragmentation.
In general for $N \geq 2$ we observe an algebraic growth of the entanglement entropy with time, $S_{L/2}(t) \sim t^{\beta_N}$, as shown in Fig.~\ref{fig:dynamics}a.
For the growth exponent $\beta_N$, we find $\beta_2 = 0.246 \pm 0.002$ for $N=2$, consistent with the expected value of $\beta_2=1/4$ arising from the EW universality class.
By contrast, for $N \geq 3$ we observe slower growth of the entanglement with $\beta_3 = 0.206 \pm 0.004$, comparable to the growth exponent found in the colored dimer deposition-evaporation model.

We assume FV scaling for the entanglement growth, yielding $S_{L/2}(t) \sim L^{\alpha} f(tL^{-z})$, to first extract the dynamical critical exponent $z$ independent of $\alpha$ by a scaling collapse analysis for an ansatz $S_{L/2}(t) = S_{L/2}(t \rightarrow \infty) f(tL^{-z})$, shown in Fig.~\ref{fig:dynamics}b.
For $N=2$, we observe diffusive dynamics with $z = 2.00 \pm 0.02$ as expected.
With more colors, we instead find subdiffusive behavior, with $N=3$ giving $z = 2.65 \pm 0.06$.
This value is consistent with the dynamical exponent in the Fredkin chain~\cite{Singh_2021, Chen_2017, Adhikari_2021} and dimer models~\cite{Koduvely_1998, Menon_1995, Menon_1997, Barma_1994}.

\begin{figure}[!t]
    \centering
    \includegraphics[width=\columnwidth]{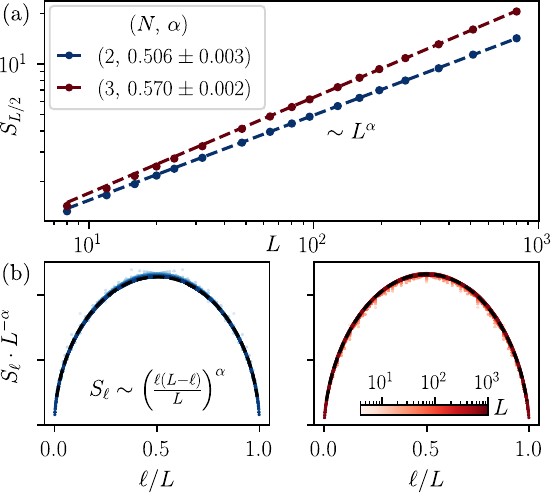}
    \caption{
    (a)
    Algebraic scaling of the steady-state half-chain entanglement entropy $S_{L/2} \sim L^\alpha$ for $N=2,3$ colors.
    With $N=2$ (blue), we find $\alpha = 0.506 \pm 0.003$, while for $N=3$ (red) we instead find $\alpha = 0.570 \pm 0.002$.
    (b)
    Data collapse of the entanglement profile $S_\ell$ confirms scaling of the form $S_\ell \sim \left(\frac{\ell(L-\ell)}{L}\right)^\alpha$ (black dashed lines).
    }
    \label{fig:SS_results}
\end{figure}

\paragraph{Steady-state entanglement scaling.} --- 
After a time $t \sim L^z$, the average entanglement entropy saturates to a sub-extensive value which scales algebraically in system size as $S_\ell(t \rightarrow \infty) \sim \left(\frac{\ell(L-\ell)}{L}\right)^{\alpha_N}$.
In Fig.~\ref{fig:SS_results}, we show the scaling of $S_{L/2}$ with respect to $L$ as well as the data collapse of the full entanglement profile $S_\ell$.
For $N=2$, we find $\alpha_2 = 0.506 \pm 0.003$, consistent with the expected EW universality.
With more colors, $N \geq 3$, we observe a slightly larger exponent $\alpha_3 = 0.570 \pm 0.002$.
Both values of $\alpha_N$ are consistent with that predicted by the dynamical scaling relation $\alpha = z \beta$.
Nonetheless, it is surprising to find $\alpha_3 > \tfrac12$, as one might expect $\alpha=\tfrac12$ from previous numerical studies and the height distribution arising in the Dyck path ensemble.

The algebraic scaling of entanglement found here is \emph{robust} so long as the color symmetry is preserved.
For example, dimerization of the measurement probability is a relevant perturbation in the $N=1$ limit, driving the system into one of two topologically distinct area-law phases.
In the colored loop ensemble with $N \geq 2$, however, such dimerization fails to alleviate the kinetic constraints, instead it only renormalizes the effective diffusion constant (or timescale for $N \geq 3$).


\paragraph{Discussion} --- 
We have introduced a non-reciprocal quantum circuit realizing the colored Temperley-Lieb algebra.
By varying the number of colors, we have shown how symmetries and kinetic constraints of the classical dynamics are imprinted upon the quantum chain, leading to robust power-law entanglement scaling and distinct dynamical exponents.
Similar ideas can also be implemented in purely unitary random circuits~\cite{Han_2024}, where kinetic constraints modify ergodicity and thermalization time scales.
Spacetime locality of the interaction implies that the classical ``control'' chain can be promoted to a quantum system, yielding a strictly quantum circuit which can be compared to a corresponding Hamiltonian dynamics.
For example, with $N=2$ we may promote the Potts spins to a chain of qubits where $\sigma_l^z$ encodes the ``color''.
Then projective measurement of $\sigma_l^z \sigma_{l+1}^z$ determines whether or not two adjacent spins have the same color.
More generally, we may consider coupling a stochastic classical cellular automaton on the control chain to the measurement-dynamics on the quantum chain, yielding a broader family of entanglement dynamics and related surface-growth models.
The power-law entanglement scaling we observed here was protected by the fact that the local color configuration on the control chain did not provide direct access to the local height field, preventing us from steering toward area-law or volume-law entanglement.
By contrast, a circuit using the Potts spins to encode left and right endpoints of Majorana loops would permit more precise tuning of transition rates in the surface growth model.
The non-reciprocal architecture we introduce here opens a new avenue for realizing a rich breadth of entanglement dynamics in quantum circuits.

\begin{acknowledgments}
K. K. was primarily supported by an NSF Graduate Fellowship under Grant No. DGE 2146752. K. K. and J. E. M. acknowledge support from the U.S. Department of Energy, Office of Science, National Quantum Information Science Research Centers, Quantum Science Center. J. E. M. acknowledges support from a Simons Investigatorship.
M. B. acknowledges support from the Deutsche Forschungsgemeinschaft (DFG, German Research Foundation) under Germany’s Excellence Strategy Cluster of Excellence Matter and Light for Quantum Computing (ML4Q) EXC 2004/1 390534769, and by the DFG Collaborative Research Center (CRC) 183 Project No. 277101999 -project B02.

\end{acknowledgments}

\bibliography{draft}

\end{document}